\documentclass[aps, superscriptaddress, amsmath, amssymb, reprint]{revtex4-1}
\usepackage{graphicx}
\usepackage{dcolumn}
\usepackage{bm}
\usepackage{tabularx}

\usepackage[utf8]{inputenc}
\usepackage[T1]{fontenc}
\usepackage{mathptmx}
\usepackage{color, soul}
\usepackage{siunitx}
\usepackage{soul,xcolor}
\usepackage{comment}
\usepackage[normalem]{ulem}
\graphicspath{ {Figures/} }

\begin{document}

\title{Observing unconventional superconductivity via kinetic inductance in Weyl semimetal MoTe$_2$}

\author{Mary Kreidel}
\affiliation{Department of Applied Physics, School of Engineering and
Applied Sciences, Harvard University, Cambridge, MA 02138}
\author{Julian Ingham}
\affiliation{Department of Physics, Columbia University, New York, NY 10027, USA}
\author{Xuanjing Chu}
\affiliation{Department of Applied Physics and Applied Mathematics, Columbia University,
New York, NY 10027, USA}
\author{Jesse Balgley}
\affiliation{Department of Mechanical Engineering, Columbia University, New York, NY 10027, USA}
\author{Ted S. Chung}
\affiliation{Department of Mechanical Engineering, Columbia University, New York, NY 10027, USA}
\author{Abhinandan Antony}
  \affiliation{Department of Mechanical Engineering, Columbia University, New York, NY 10027, USA}
\author{Nishchhal Verma}
\affiliation{Department of Physics, Columbia University, New York, NY 10027, USA}
\author{Luke N.~Holtzman}
\affiliation{Department of Applied Physics and Applied Mathematics, Columbia University,
New York, NY 10027, USA}
\author{Katayun Barmak}
\affiliation{Department of Applied Physics and Applied Mathematics, Columbia University,
New York, NY 10027, USA}
\author{Raquel Queiroz}
\affiliation{Department of Physics, Columbia University, New York, NY 10027, USA}
\author{James Hone}
\affiliation{Department of Mechanical Engineering, Columbia University, New York, NY 10027, USA}
\author{Robert M.~Westervelt}
\affiliation{Department of Physics, School of Engineering and
Applied Sciences, Harvard University, Cambridge, MA 02138}
\affiliation{Department of Applied Physics, School of Engineering and
Applied Sciences, Harvard University, Cambridge, MA 02138}
\author{Kin Chung Fong}%
\email{present address: k.fong@northeastern.edu}
\affiliation{RTX BBN Technologies, Quantum Engineering and Computing Group, Cambridge, MA 02138, USA}
\date{\today}

\begin{abstract}
Identifying the pairing symmetry of unconventional superconductors plays an essential role in the ongoing quest to understand correlated electronic matter. A long-standing approach is to study the temperature dependence of the London penetration depth $\lambda$ for evidence of nodal points where the superconducting gap vanishes. However, experimental reports can be ambiguous due to the requisite low-temperature resolution, and the similarity in signatures of nodal quasiparticles and impurity states. Here we study the pairing symmetry of Weyl semimetal $T_d$-MoTe$_2$, where previous measurements of $\lambda$ have yielded conflicting results. We utilize a novel technique based on a microwave resontor to measure the kinetic inductance of MoTe$_2$, which is directly related to $\lambda$. The high precision of this technique allows us to observe power-law temperature dependence of $\lambda$, and to measure the anomalous nonlinear Meissner effect --- the current dependence of $\lambda$ arising from nodal quasiparticles. Together, these measurements provide smoking gun signatures of nodal superconductivity. 
\end{abstract}
 
\maketitle

Elucidating the nature of unconventional superconductivity is a central focus of modern condensed matter physics. This effort is crucial for discovering materials with high critical temperatures, and identifying platforms for topological quantum computing \cite{sato2017topological}. A hallmark of unconventional superconductivity is a vanishing gap at nodes on the Fermi surface, which allow Cooper pairs to avoid each other and minimize Coulomb repulsion \cite{Sigrist.1991}. 
The vanishing gap leads to unpaired quasiparticles, and in turn a nonzero density of states (DOS) at low energy that is absent in fully gapped superconductors. The difference in low-energy DOS produces different thermal excitations in nodal and fully gapped superconductors, and therefore different temperature dependence in, for instance, their transport of heat and sound \cite{arfi1988thermal, hayes2025field, luthi2005ultrasonics}. 
While nodal superconductivity has been observed in high-$T_c$ copper oxides \cite{Annett.1991, Hardy.1993}, superfluid $^3\text{He}$ \cite{leggett1975theoretical}, and heavy fermion compounds \cite{Norman.Norman.2011}, its existence has been difficult to establish in many other systems due to experimental challenges. In particular, conventional techniques lack the sensitivity for the study of van der Waals (vdW) superconductors --- an emerging platform for the study of how rich correlated physics can coexist with novel superconductivity \cite{Wang.2020, Kim.2024, Banerjee.Kim.2024, tanaka2025superfluid}.

An important method for identifying nodal superconductivity is the measurement of the London penetration depth $\lambda$: the distance across which external magnetic fields are expelled in accordance with the Meissner effect \cite{Prozorov.2006}. Since thermal activation of low-energy states increases $\lambda$, its temperature-dependence reflects the DOS: $\lambda$ saturates exponentially as $T\rightarrow 0$ for clean fully gapped superconductors, but shows power-law behavior in nodal superconductors. Measurements of $\lambda$ have played a key role in identifying the pairing symmetry of many nodal superconductors \cite{Annett.1991, Hardy.1993, Hashimoto.2012, Gross.1986}. However, in ``sign-changing'' fully gapped superconductors --- for which the gap changes sign on different Fermi sheets --- impurities can introduce low-energy DOS which also lead to power-law behavior in $\lambda$ \cite{Bang.2009,Vorontsov.2009}, complicating the identification of nodal superconductivity.

Yet, there remains a fundamental difference between nodal quasiparticles and impurity-induced DOS: since the latter are typically localized, they do not move in response to an applied current. Therefore measuring the dependence of $\lambda$ on applied magnetic field or current $I$ --- known as the nonlinear Meissner effect --- can distinguish these two types of low-energy states \cite{Yip.1992, Xu.1995, Stojkovic.1995, zhuravel2013imaging, Sauls.2022}. In particular, nodal superconductors should display an anomalous nonlinear Meissner effect (ANLME) \cite{Yip.1992} that is distinct from the behavior of nodeless superconductors even in the presence of impurities. However, observation of this effect has been experimentally challenging: for low-$T_c$ materials, it requires measurement of very small changes in $\lambda$ at $T\ll T_c$; in high-$T_c$ cuprates, the ANLME has been difficult to distinguish from effects of vortex flow \cite{poole2014superconductivity}.

Here we overcome these challenges by employing a resonator technique~\cite{Mary.KC.2024} to measure parts-per-million changes in $\lambda$ with temperature and current. We apply this technique to transition metal dichalcogenide \mbox{$T_d$-MoTe$_2$}, a vdW type-II Weyl semimetal \cite{Qi.2016, Deng.2016,Tamai.2016,Jiang.2017,Rhodes.2017,Sun.2015}. The non-trivial topology of the normal state makes MoTe$_2$ a fascinating playground to explore the interplay of Weyl fermions and electronic correlations, implicated by theoretical studies in the possible formation of unconventional gap functions \cite{Li.2018bwq,Hosur.2014,Yan.2020}. A number of interesting features of the superconducting state have already been observed \cite{Jindal.2023,du2024superconducting}, including novel edge supercurrent \cite{Wang.2020}, and a sharp increase in $T_c$ in the monolayer limit \cite{Rhodes.2021}. Previous works have claimed either conventional $s_{++}$ pairing \cite{Piva.2023}, or sign-changing $s_{\pm}$ pairing \cite{Luo.2020, Guguchia.2017}. Despite this prevailing view, recent measurements of Josephson tunneling between MoTe$_2$ and Nb have provided tentative indications that the superconducting state in MoTe$_2$ might not be $s$-wave  \cite{Kim.2024}.

\begin{figure*}
\centering
\includegraphics[width=1.2\columnwidth]{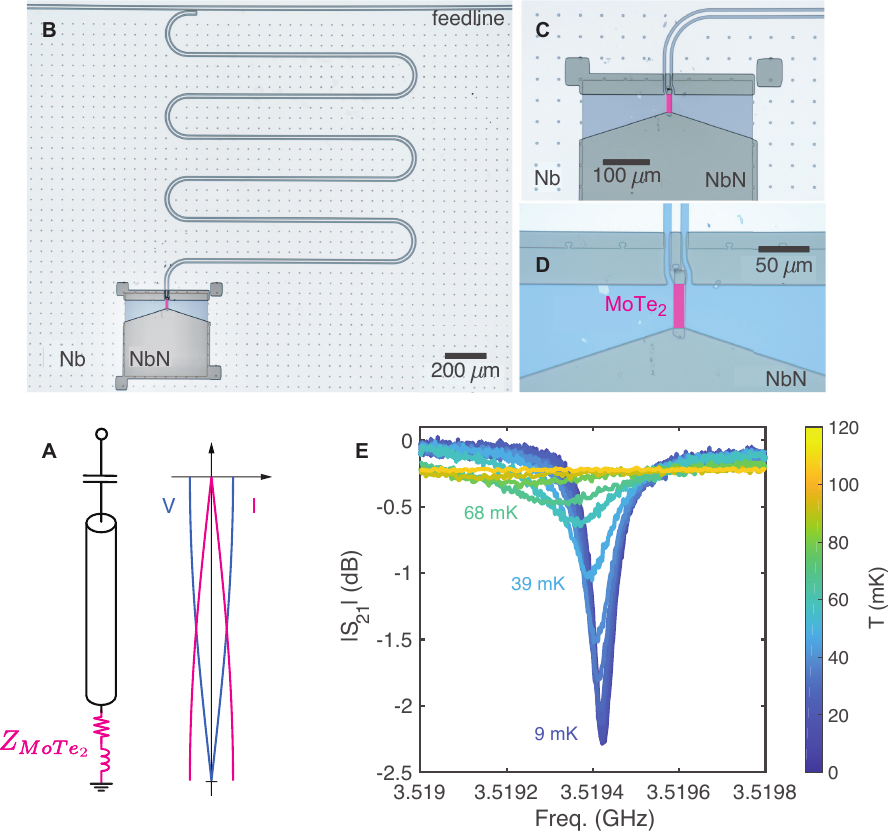}
\caption{\textbf{Measuring the kinetic inductance of superconducting MoTe$_2$.}  \textbf{(A)} Schematic of the quarter-wave resonator experiment. The transmission-line --- represented by a coaxial cable (cylinder) --- is terminated to the ground through the MoTe$_2$ sample. When the magnitude of the complex impedance $Z_{MoTe_2}$ --- represented by the resistor and inductor in series --- is small compared to the characteristic impedance of the transmission line, the transmission line and sample form a quarter-wave resonance with the sample located at the current antinode. \textbf{(B-D)} Optical micrographs of the resonator. \textbf{(B)} The meandering coplanar-waveguide made of niobium is capacitively coupled to the horizontal feedline for transmission measurement at microwave frequencies. \textbf{(C)} The resonator terminates to the niobium ground plane through \textbf{(D)} an exfoliated flake of MoTe$_2$ (pink rectangle), connected galvanically with sputtered niobium nitride. The MoTe$_2$ flake, shown in pink false-coloring, is oriented such that the long edge of the pink rectangle is along the $a$-axis of the MoTe$_2$ crystal. \textbf{(E)} Measured $|S_{21}|$ of the MoTe$_2$ hybrid resonator at various temperatures.}
\label{fig:expt}
\end{figure*}

To obtain $\lambda$, we measure kinetic inductance $L_K$, which arises from the inertia of the superfluid in response to oscillating currents. Kinetic inductance has recently been used to study the penetration depth of other vdW superconductors \cite{Mary.KC.2024, Banerjee.Kim.2024, tanaka2025superfluid}; in the local limit, $L_K$ is proportional to $\lambda^2$,
\begin{equation}
     L_K = \frac{\mu_0 l}{2wt} \lambda ^2 , \label{eq:London}
\end{equation}
where $\mu_0$ is the vacuum permeability, and $w$, $l$ and $t$ are the width, length, and thickness of the superconductor. In the resonator method, we connect the sample between a superconducting coplanar waveguide and the ground plane, schematically shown in Fig.~1A. When the sample impedance $|Z|$ is much smaller than the characteristic impedance ($50~\Omega$) of the waveguide, the sample and waveguide form a quarter-wave resonator. At the current antinode, an increase in $L_K$ with temperature or current will induce a proportional decrease in the resonance frequency $f_\mathrm{res}$:  
\begin{equation}
    \frac{\delta f_{\rm res}}{f_{\rm res}} = -\frac{8}{\pi^2}\frac{\delta L_K}{L_0}\rm{,}\label{eqn:FreqShift}
\end{equation} where $L_0 \simeq 2.8$~nH is the equivalent inductance of the quarter-wave resonator. The high-$Q$ resonator enables precise tracking of $f_{\mathrm{res}}$ to achieve the required high sensitivity to measure the changes of $\lambda$ in MoTe$_2$. We have previously validated this approach on Al and NbSe$_2$ samples \cite{Mary.KC.2024}.

\begin{figure*}[t!]
    \centering
    \includegraphics[width=\textwidth]{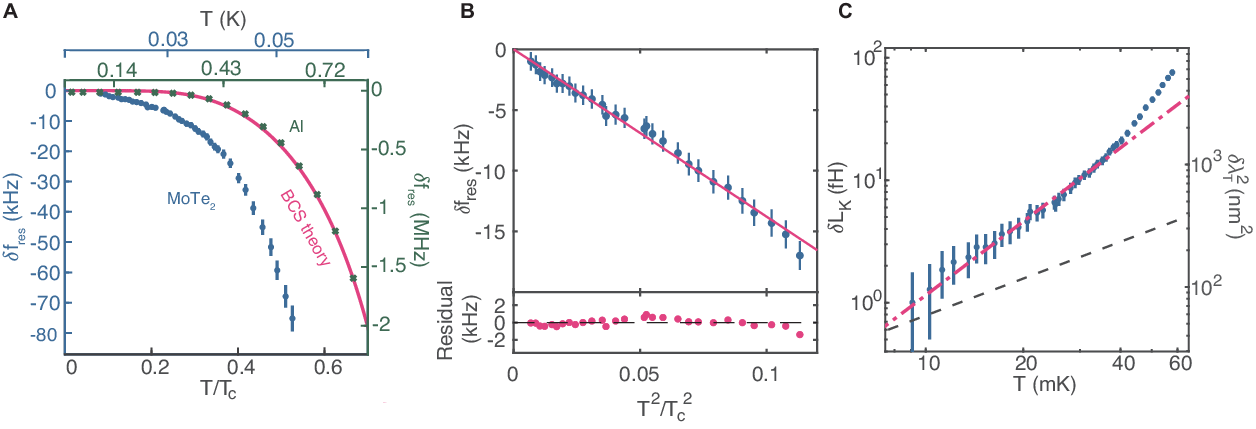} 
    \caption{\textbf{Resonance frequency shift under thermal excitation.} \textbf{(A)} Measured frequency shift, $\delta f_{res}$, versus temperature normalized to the superconducting transition temperature; $\delta f_{res}$ is referenced to 3.519420 and 3.258497 GHz for MoTe$_2$ and Al, respectively. As $T\rightarrow 0$, the frequency rises as the kinetic inductance of MoTe$_2$ (blue dots) decreases. In contrast to the aluminum data (green crosses), which saturate at around $0.3T_c$ in accordance with BCS theory (pink solid line). The double $x$-axes at the top of the plot denote the absolute temperatures for the MoTe$_2$ (blue) and Al (green) data. \textbf{(B)} The normalized frequency shift is plotted against $(T/T_c)^2$ and fit to a linear trend (solid pink line). Exponential or linear scaling data would feature a systematic drift away from the best fit as $T\rightarrow0$, yet the residual plot (bottom) shows no such trend. The scaling behavior is precisely characterized by plotting $\delta L_K$ (left axis) or equivalently $\delta \lambda^2_T$ (right axis) on a log-log plot, with quadratic scaling (pink dashed line) contrasted with linear scaling (black dashed line); the slope of best fit is $n=2.12 \pm 0.15$. The change in kinetic inductance along the $a$-axis of the MoTe$_2$ flake is on the order of a few fH in this low-temperature regime. The magnitude of $\delta L_K$ and the error bar underscores the stringent sensitivity required and achieved in this experiment.}
    \label{fig:FreqVsT}
\end{figure*}

Figs.~1B--D show optical microscope images of one device used in this study, with the MoTe$_2$ sample false-colored in pink. Each sample is fabricated by first patterning the resonator on a Si wafer, then placing an exfoliated MoTe$_2$ flake onto the end. Five samples with varying dimensions (Table~\ref{tab:SampleDimensions}) were studied, all in the thickness regime (82-289 nm) where bulk superconductivity is expected. Electrical contact between the MoTe$_2$ and the waveguide and ground plane is made by e-beam lithography and deposition of niobium nitride by reactive-ion sputtering. As a control, we terminate an identical resonator with a thin aluminum wire (see Methods for fabrication details).

To effectively cool our samples to low temperatures, we package the devices in copper enclosures that are designed to suppress electromagnetic noise and provide good thermalization to the dilution refrigerator. External magnetic fields are shielded by high-permeability mu-metal and superconducting enclosures. The probe signal is delivered through a coaxial cable that is attenuated at each cryostat stage to minimize thermal noise. After interrogating the resonators through coupling capacitors, the probe signal propagates through an isolator before amplification by a cryogenic low-noise amplifier. We characterize the resonators using the scattering parameter, $S_{21}$ (see Methods for details).

Fig.~1E shows representative data obtained using this technique. The resonance is seen as a sharp dip in the scattering parameter $|S_{21}|$ at the base temperature of 9~mK. As the temperature increases, minute decreases in $f_{\text{res}}$ due to increasing $L_K$ are clearly observable, along with broadening of the resonance due to dissipation from thermally activated quasiparticles in the MoTe$_2$ sample. The data are taken at 0.2~fW at the resonator input --- a low power corresponding to as few as 40 photons --- to mitigate nonlinear effect. For each dataset, circle fitting (see Methods) is used to extract $f_{\text{res}}$ and $Q$ \cite{Probst.2015}.  

We first investigate the temperature scaling of $\lambda^2$ by examining the shift from its zero-temperature value, $\delta\lambda^2_T \equiv \lambda^2(T)-\lambda^2(T=0)$.  We plot the $f_{\rm res}$ data of the MoTe$_2$ resonator and aluminum control against normalized temperature ($T/T_c$) in Fig.~2A. For Al, $f_{\mathrm{res}}$ saturates to a constant value for $T\rightarrow 0$ as expected for a fully gapped, conventional $s$-wave superconductor. The temperature dependence of $f_{\rm res}$, and hence of both $L_K$ and $\lambda^2$ agrees well with BCS theory (pink line), with fit parameters yielding $T_{c} = 1.195 \pm 0.001$~K, and normal-state sheet resistance $109 \pm 26$~m$\Omega$ matching DC electrical transport measurements. By contrast, for MoTe$_2$, $f_{\rm res}$ does not saturate at low $T$, but continues to rise with decreasing $T$ down to 9~mK.

\begin{figure*}[t]
\centering
\includegraphics[width=2\columnwidth]
{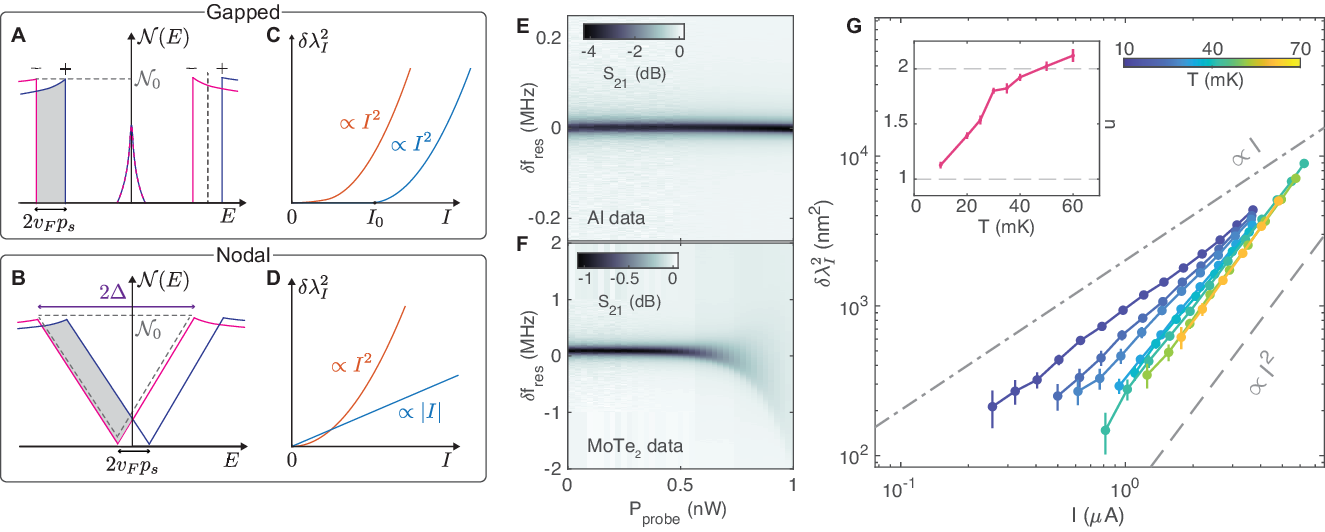} 
\vspace{-0.3cm}
\caption{\textbf{Anomalous nonlinear Meissner effect in MoTe$_2$.} 
Density of states (DOS) under an applied current for \textbf{(A)} gapped and \textbf{(B)} nodal superconductors. The applied current shifts the quasiparticle momenta by an amount $p_s$; current dependence of $\lambda^2$ results from a difference in the DOS for forward ($+$) and backward ($-$) moving quasiparticles (shaded region). Impurity-induced DOS near $E\approx 0$ for unconventional, fully gapped superconductors do not shift in energy when the momenta shift by $p_s$, and so do not produce a contribution to the shaded region, whereas nodal quasiparticles with a linear dispersion produce a contribution to the shaded region inside the gap $\propto$~$p_s|p_s|$. The shaded areas are given approximately by a rectangle, and difference of two triangles, in the two respective cases. \textbf{(C)} The current dependence of $\lambda^2$ for a fully gapped superconductor at zero (blue) and finite (red) temperature. \textbf{(D)} The current dependence of the superfluid density for a nodal superconductor at zero (blue) and finite (red) temperature. In the nodal case, a current dependent shift $\propto$~$|I|$ appears at small $I$. 
 \textbf{(E)} Measured frequency change of an Al hybrid resonator compared to \textbf{(F)} that of an MoTe$_2$ hybrid sample resonator at the same range of input powers, taken at $T \simeq 10$~mK. 
\textbf{(G)} Measured $\delta \lambda_I^2$ as a function of current across various temperatures, shown on a log-log scale to illustrate power-law behavior. The power law $n$ of best fit as a function of temperature is plotted in inset.}
\label{fig:NLME}
\vspace{-0.34cm}
\end{figure*}

To quantify the scaling of $\delta\lambda^2_T$ for MoTe$_2$, we first must extrapolate the data to estimate $f_{\rm{res}}(T=0)$. To do this, we plot $\delta f_{res}$ against $(T/T_c)^n$ for different $n$, and find that the data are best described by quadratic scaling  (Fig.~2B and Fig.~S8, further details in SI Section C). Using Eqs.~(1) and (2), we then plot $\delta L_K$ and $\delta\lambda_T^2$ versus $T$ on a log-log scale. Fig.~2C displays fH-scale variations in $L_K$, underscoring the stringent sensitivity required and achieved in this experiment. Across 9~mK $<T<$ 40~mK, the data converge self-consistently to a line, signifying power-law behavior $\delta\lambda^2 \propto T^n$; the best fit yields $n = 2.12 \pm 0.15$. We perform fine temperature scans of four MoTe$_2$ samples that reproduce the $T^2$ scaling (see Table~\ref{tab:TdepPower} in Supplementary Information). 

To understand the consequences for the pairing symmetry in MoTe$_2$, we first note that in nodal superconductors, the presence of low-energy DOS means that $\lambda^2_{T}$ shows power-law temperature scaling. For gaps with line nodes, $\sim T$ and $\sim T^2$ scaling are expected in the clean and moderate impurity scattering limits respectively --- both of which have been seen in the high-$T_c$ cuprates \cite{Gross.1986,Hirschfeld.1993} --- while gaps with point nodes give $\sim T^2$ in the clean limit. Our data are therefore consistent with either an impure line-nodal superconductor, or a clean point-nodal superconductor.

By comparison, conventional fully gapped superconductors have no states inside the superconducting gap, $\Delta$ --- even in the presence of non-magnetic impurities, as stipulated by Anderson's theorem \cite{anderson1959theory} --- so that $\delta \lambda_T^2 \sim e^{-\Delta/k_BT}$. However, fully gapped superconductors with sign changing $s_{\pm}$ pairing can acquire in-gap states from impurity-induced interband scattering. The resulting low-energy states give rise to $\delta \lambda_T^2 \sim T^n$ scaling with $n\approx 3$ approaching $2$ for strong disorder, as observed in certain iron-based superconductors \cite{Bang.2009,Vorontsov.2009}. Thus, observation of power-law scaling in $\delta\lambda^2_{T}$ cannot conclusively discriminate between nodal superconductivity, and unconventional $s$-wave pairing in the presence of impurities.

As discussed above, the ambiguity between nodal quasiparticles and impurity-induced low-energy states can be resolved by measuring the scaling of $\lambda^2$ with applied current $I$. We define its change relative to $I=0$ as $\delta\lambda^2_I \equiv \lambda^2(I,T)-\lambda^2(I=0,T)$. Under an applied current $I$, the quasiparticle momenta are shifted by $p_s \propto I$; the current-velocity constitutive relation is $j_s = (1/\mu_0e\lambda^2)p_s$, with $e$ the electron charge. A linear Meissner effect --- for which $j_s$ depends linearly on $p_s$ and hence $I$ --- means that $\lambda^2$ is independent of $I$. On the other hand, a \textit{nonlinear} Meissner effect --- for which $j_s$ depends nonlinearly on $p_s$ --- means that $\lambda^2$ depends on $I$. 

Quantitatively, we can calculate $j_s$ using \cite{Xu.1995}
\begin{gather}
j_{s}=2e v_F \int f_T(E)\left(\mathcal{N}\left(E, p+p_{s}\right)-\mathcal{N}\left(E, p-p_{s}\right)\right) \ dE  
\label{eqn:qpCurrent}
\end{gather} where $v_F$ is the Fermi velocity, $f_T(E)$ the Fermi-Dirac function, and $\mathcal{N}\left(E, p \pm p_{s}\right)$ is the DOS at energy $E$ for forward and backward moving quasiparticles with respect to the supercurrent (see Methods). The shift in the quasiparticle energy $\pm v_F p_s$ is known as a Doppler shift, which leads to a number of other phenomena in nodal superconductors \cite{matsuda2006nodal}. The shaded areas in Fig.~3A and 3B represent the integrand in Eq.~\ref{eqn:qpCurrent} for fully gapped and nodal superconductors, respectively. The distinct nonlinear behavior is a direct consequence of the difference in their DOS.

For a fully gapped superconductor and small $p_s$, the shaded area is $2\mathcal{N}_0 v_Fp_s $, where $\mathcal{N}_0$ is the DOS in the normal state. Therefore, $j_s = 2e\mathcal{N}_0v_F^2p_s$. The constitutive relation implies $\lambda^2$ is independent of $I$, i.e. the Meissner effect is linear. Nonlinearity emerges for a Doppler shift on the order of the gap energy when $I$ is comparable to the critical current, i.e. $I\sim I_0$, leading to $\delta \lambda_I^2 \propto I^2$. At finite temperature, thermally excited quasiparticles contribute an exponentially-suppressed current shift even at small $p_s$, resulting $\delta \lambda_I^2 \propto a(T)I^2$, with $a(T)\propto e^{-\Delta/k_BT}$; the current dependence of conventional superconductors is summarized in Fig. 3C. 

By contrast, the low-energy states from nodal quasiparticles yield a dependence of the shaded area on $p_s$ even for small $p_s$ and zero temperature --- that is $2\mathcal{N}_0 v_Fp_s(1-v_F|p_s|/2\Delta)$ (Fig.  3B). In addition to the $p_s$-independent contribution to $\lambda^2$, there is a nonlinear term proportional to $|p_s|$, and hence $\delta\lambda_I^2 \propto |I|$; this is the ANLME \cite{Sauls.2022}. At elevated temperatures, this crosses over to $\delta\lambda_I^2 \propto I^2$ dependence \cite{Xu.1995}. Fig. 3D summarizes the ANLME in nodal superconductors.

Crucially, low-energy impurity states in fully gapped $s_\pm$ superconductors --- depicted as a peak near $E\approx 0$ --- can contribute to the temperature dependence of $\lambda^2$. These states are typically localized, so do not experience a Doppler shift; a temperature or current dependent shift in the penetration depth requires impurity states hybridize to form an impurity band, fully closing the gap and producing gapless superconductivity. Yet these extended states lack the linear dispersion of those near nodal points, and so only produce a $\sim I^2$ nonlinearity, as in the conventional case. Hence, impurity states do not contribute a $\sim |I|$ dependence to $\delta \lambda^2$ (see Methods). Therefore, the ANLME sharply discriminates between nodal and unconventional nodeless superconductivity, allowing us to precisely observe nodal pairing in MoTe$_2$.

We measure $\delta\lambda^2_I$ by tracking the shift in $f_{\mathrm{res}}$ with probe power $P_{\rm probe}$. Fig.~3E and 3F display $S_{21}$ vs frequency and $P_{\rm probe}$ for Al and MoTe$_2$. While $f_{\mathrm{res}}$ remains constant for the aluminum sample,  it decreases significantly with increasing $P_{\rm probe}$ for the MoTe$_2$ sample. This evidences a nonlinear response that is absent in the fully gapped aluminum sample at low $P_{\rm probe}$.

To quantitatively determine $\delta\lambda^2_I$, we use Input-Output theory (Methods) with the extracted quality factor of the resonator to obtain $I$ as a function of $P_{\rm probe}$. Figure 3G plots $\delta\lambda_I^2$ against $I$ at various temperatures. Remarkably, at 9~mK  $\delta\lambda_I^2$ exhibits power-law scaling $I^{n}$ with a best fit power law $n=1.12\pm0.02$, demonstrating the ANLME. Five different samples were used for current dependent measurements, and the linear scaling of each is presented in Fig. 7 of the Supplementary Information. We further investigate the temperature dependence of the ANLME in two samples. As a function of temperature, we find $\delta\lambda^2_I\sim I^n$ scaling with $n$ steadily increasing to $n=2.14\pm0.02$ at 60 mK (Fig.~3G, inset). At higher temperatures, the quality factor of the resonator drops below $10^4$, degrading the sensitivity of our measurement and restraining us from determining accurately the exponent. In the second sample, we find that $n$ increases from $1.55\pm 0.04$ to $2.18\pm 0.08$ (SI). The difference between the exponents in the two devices at 9 mK could arise due to anisotropy in the Fermi velocity --- since the second sample was etched along the crystal $b$-axis rather than cleaved along the $a$-axis --- or to the more rapid degradation of the $Q$ factor we observe in the second device (SI); we leave a more rigorous exploration of the temperature dependence of $\delta\lambda^2_I$ to future work. 

Our observed temperature dependence of the ANLME agrees with theory \cite{Xu.1995}, which predicts the scaling of $\delta\lambda_I^2$ to evolve from $I$ to $I^2$ as temperature increases. The temperature at which this crossover is predicted to occur is $T_{c}\cdot\left(H_{c 1} / H_{c 2}\right) \simeq 10$~mK for MoTe$_2$, consistent with the temperature window across which we observe this effect. More importantly, the temperature dependence of the $I^n$ power law in our data can decisively rule out an alternative explanation of linear $I$ dependence of $\delta\lambda^2$ due to vortex flow \cite{Wilcox.2022}; such a $\propto I$ contribution grows stronger with increasing temperature, rather than evolving from $I$ to $I^2$ as demonstrated in our data. As discussed, the scaling $\delta \lambda^2_I \sim I$ is inconsistent with a fully gapped superconductor, with or without impurities, and with or without a sign-changing $s_\pm$ gap. Taken together, we therefore interpret the observed behavior of $\delta \lambda^2_T$ and $\delta \lambda^2_I$ in terms of the presence of nodes in the superconducting gap of $\mathrm{MoTe}_{2}$.

We now discuss superconducting states which are consistent with this phenomenology. The observed $\delta\lambda_T^2\sim T^{2}$ scaling can be explained either by a line-nodal superconductor with impurities, or a clean superconductor with point nodes. Yet the nonlinear Meissner effect is suppressed in the presence of disorder \cite{Xu.1995,Wilcox.2022}, and so we consider the former scenario unlikely. A superconductor with point nodes would naturally produce both $\delta\lambda^2_T  \sim T^2$ and the cross over from $\delta\lambda^2_I\sim I$ to $\sim I^2$ with increasing temperature. We consider symmetry arguments to help identify candidate gap structures; $\mathrm{MoTe}_{2}$ forms an orthorhombic crystal structure with space group $P m n 2_{1}$ and point group symmetry $C_{2 v}$ \cite{Qi.2016}. Possible superconducting gaps are classified by their irreducible representation of $C_{2 v}$, namely $A_{1}, A_{2}, B_{1}, B_{2}$ depending on their transformations under $C_{2 z}$ and mirror symmetries \cite{Bradley.1972}. A superconducting gap with point nodes requires an unconventional triplet state, which is odd under mirror symmetry $\Delta(k) \sim\left(k_{x} s_{x}+k_{y} s_{y}\right) i s_{y}$ --- i.e. the $A_{2}$ representation --- which possesses point nodes at $k=0$. The presence of spin-orbit coupling and broken inversion symmetry results in singlet-triplet mixing, but this triplet state would develop a singlet component $\sim\left(k_{x} k_{y}\right) i s_{y}$ which does not remove the point nodes. Prior studies of pairing in topological bands have proposed that such superconducting phases may arise naturally due to Dirac/Weyl physics \cite{Li.2018bwq,Hosur.2014,Yan.2020}; we leave a more thorough microscopic analysis to future work. We note interestingly that recent ARPES data have shown evidence for nodal superconductivity in Weyl semimetal PtBi$_2$, motivating further theoretical scrutiny of nodal pairing in Weyl semimetals broadly \cite{changdar2025topological}.

Our results motivate a renewed scrutiny of the pairing symmetry in MoTe$_2$, and transition metal dichalcogenide superconductors more generally. Precision measurements of the thermal conductivity and specific heat could also investigate the presence of power-law temperature dependencies --- though as demonstrated by our present work, it is imperative that these measurements extend to very low temperatures to reliably distinguish power-law behavior from exponential saturation. Scanning tunneling microscopy might also be used to measure the in-gap density of states, though similar challenges arise due to the low $T_{c}$ of MoTe$_2$ and therefore requisite high resolution. Josephson tunneling could also be employed to probe sign changes of the superconducting gap --- as in experiments which have produced sharp illustration of $d_{x^{2}-y^{2}}$ gap symmetry in the cuprates  \cite{Wollman.1993,Mathai.1995,Brawner.1994}. Alternative probes of the ANLME include magnetic torque \cite{Sauls.2022}, and frequency intermodulation \cite{Dahm.1997}, suggesting additional approaches to verifying our results. 

The results presented herein provide smoking-gun evidence of nodal superconductivity in $\mathrm{MoTe}_{2}$ via the ANLME, in conjunction with temperature-dependent penetration depth measurements. Furthermore, we present a hybrid high-$Q$ technique as a prototype to motivate further study of superconductivity in topological or strongly-correlated materials --- offering a powerful tool in determining superconducting pairing symmetries.

\section*{Acknowledgements}
We thank M.~Randeria, J.~A.~Sauls, and B.-I.~Wu for fruitful discussions. M.~K. acknowledges support from the STC Center for Integrated Quantum Materials, NSF Grant No.~DMR-1231319. K.~C.~F., X.~C., and J.~H.~acknowledge  support from the Army Research Office under Contract W911NF-22-C-0021. 
R.~Q.~acknowledges support from the NSF MRSEC program at Columbia University through the Center for Precision-Assembled Quantum Materials (DMR-2011738).
The synthesis of MoTe$_2$ (L.H.~and K.B.) and theory analysis (R.Q.) were supported by the Columbia University Materials Science and Engineering Research Center (MRSEC) through NSF grant DMR-2011738.
The authors acknowledge the use of facilities and instrumentation supported by NSF through the Columbia University, Columbia Nano Initiative, and the Materials Research Science and Engineering Center DMR-2011738. J.B.~acknowledges support from the Army Research Office under Grant Number W911NF-24-1-0133. The views and conclusions contained in this document are those of the authors and should not be interpreted as representing the official policies, either expressed or implied, of the Army Research Office or the U.S. Government. The U.S. Government is authorized to reproduce and distribute reprints for Government purposes notwithstanding any copyright notation herein.

\section*{Methods}

\subsection*{Measurements}

\begin{figure}[tbh!]
\centering
\includegraphics[width=0.45\columnwidth]{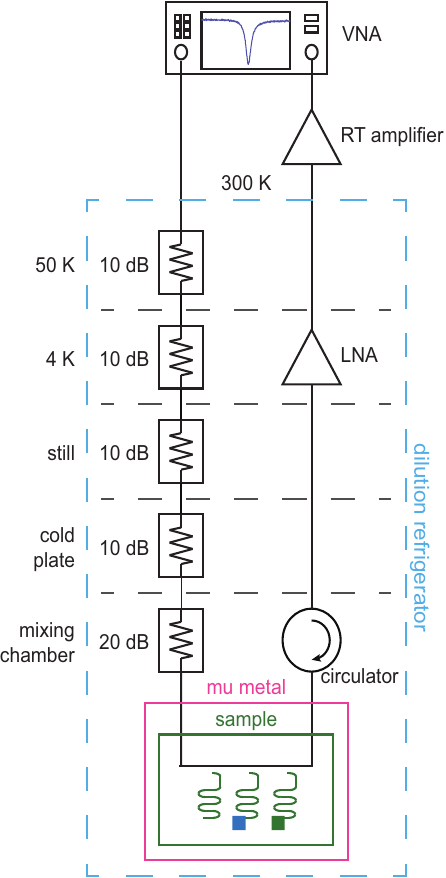} 
\caption{\textbf{Cryogenic microwave transmission measurement setup.} Schematic of the vector network analyzer (VNA)-based measurement chain used to probe superconducting resonators at millikelvin temperatures. The VNA output is attenuated by 60--80~dB using attenuators that are distributed at room temperature outside the cryostat and at various temperature stages inside the cryostat. The signal is routed through a coaxial line to the mixing chamber stage, where it excites a hanger-type resonator mounted inside a heavily shielded sample enclosure. The transmitted signal is amplified by a cryogenic low-noise amplifier (LNA) at 4~K, followed by additional room-temperature amplification, and is then returned to the VNA for readout. A circulator at the base stage isolates the sample from the noise generated by the amplifier and ensures unidirectional signal routing. The measurement chain was optimized to minimize thermal loading, eliminate electromagnetic noise, and maximize the measurement sensitivity of the resonator at powers down to the single-photon level.}
\label{fig:SetupSchematic}
\end{figure}

All measurements were performed in dilution refrigerators with a base electron temperature of 9--20~mK. Base temperature of 9~mK was attained through careful thermal anchoring, extensive electromagnetic shielding, and multi-stage filtering within copper enclosures. High-frequency transmission measurements were conducted using a vector network analyzer (VNA). The setup and circuitry are similar to that of superconducting qubit measurements and is shown schematically in Fig.~\ref{fig:SetupSchematic}. 

The VNA-generated microwave signal was attenuated by 40~dB using distributed cryogenic attenuators, followed by an additional 20~dB of attenuation just above the mixing chamber stage. The attenuated signal was coupled into the MoTe$_2$ via a quarter wave, high-Q, niobium resonator terminated by the sample, as illustrated in Fig~\ref{fig:expt}B--D. Upon interacting with the sample and resonators, the transmitted signal acquires a change in its phase and magnitude (see SI). It exited through the output port of the resonator, passed through a circulator at the mixing chamber for isolation, and was amplified in two stages: first by a cryogenic low-noise amplifier (40~dB gain) at 4~K, and subsequently by a room-temperature amplifier providing an additional 30~dB of gain, before returning to the VNA. To validate the integrity of our measurement setup and exclude artifacts, we also characterized aluminum-, NbSe$_2$, and NbN-terminated resonators in prior work \cite{Mary.KC.2024}, and comparisons to some of those results are included in the main text (Fig.~2A and \ref{fig:NLME}E).

We mount the resonators onto a cold finger of the dilution refrigerator within a copper package shielded by a mu-metal can to eliminate spurious microwave modes and stray magnetic fields. To ensure thermal equilibrium, the system was held at each temperature point for 15–20 minutes. Sample thermalization was optimized through a combination of design strategies: extensive filtering by attenuators, carefully routed and affixed wiring within the cryostat for thermal anchors, and effective electromagnetic shielding. Efficient thermalization is evident from the consistent temperature-dependent changes observed in the kinetic inductance even down to 9~mK, as well as additional DC measurements performed on normal-insulator-superconductor tunnel junction samples in the same fridge for a different research project.

Fabrication of our samples (Fig.~\ref{fig:expt}) starts with the niobium resonators on high resistivity, float-zone silicon wafers.  The niobium resonators are fabricated based on the same procedures used in superconducting qubit experiments and our previous work observing kinetic inductance of van der Waals flakes \cite{Mary.KC.2024} (lithography and plasma etching of a 200~nm niobium film sputtered onto float-zone silicon). The flakes of MoTe$_2$ are exfoliated and transferred by the dry pickup technique \cite{Dean.2010} in an inert N$_2$ environment and dropped off in the etched destination area at the end of the resonator (Fig.~1C). The crystalline $a$-axis of each flake is oriented along the long edge.  \emph{In situ} argon ion milling is used to remove surface oxide layers from the resonator and MoTe$_2$ flake, which we then connect by depositing a 3-nm-thick titanium sticking layer and reactive sputtering of 200--240 nm~of NbN \cite{Antony.2021} (Fig.~1D). NbN $T_c \simeq$14~K, higher than that of Nb (8~K). We sputter a relatively thick patch of NbN not only to ensure contact over the $\sim$100--300~nm thick MoTe$_2$, but also to reduce the contribution of kinetic inductance from NbN, which is inversely related to film thickness.

\subsection*{Nonlinear Meissner effect}

Following Ref. \cite{Xu.1995}, we write the supercurrent including the angle dependence $\theta$ on the Fermi surface, 
\begin{gather}
j_{s}=2e \int d\theta \ v_F(\theta) \nonumber\\
\times \int f_T(E)\left(\mathcal{N}\left(E, p+p_{s}\right)-\mathcal{N}\left(E, p-p_{s}\right)\right) \ dE  
\label{eqn:qpCurrent_Methods}
\end{gather}
For an isotropic Fermi velocity --- the case discussed in the main text, as reflected by Eq. \eqref{eqn:qpCurrent} --- this object is simply $2e v_F$ times an energy integral of the difference in DOS, which measures how many states contribute to the current (Figs.~3A and 3B, shaded area). At nonzero temperature $T>0$, the presence of the Fermi function $f_T(E)$ results in states with $E>0$ contributing to the integral --- physically corresponding to thermally excited quasiparticle states which reduce the supercurrent, or equivalently increase the penetration depth. We comment also that the supercurrent can be written in terms of the so-called superfluid stiffness $\mathcal{D}_s$, $j_s = (4e\mathcal{D}_s/\hbar^2 )p_s$, i.e. $\mathcal{D}_s = (\hbar^2/4\mu_0e^2)\lambda^{-2}$. The nonlinear Meissner effect can therefore equivalently be thought of as manifesting a current-dependent effective superfluid stiffness. 

For a fully gapped superconductor and small $p_s$ (Fig.~3C), the shaded area is approximately a rectangle with area $\mathcal{N}_0 v_F p_s$. Therefore $j_{s} = 2e \mathcal{N}_0 v_F^2 p_s$ such that
\begin{align}
    \lambda^{-2}= 2\mu_0 e^2 \mathcal{N}_0 v_F^2
\end{align}
is independent of $I$. For large $p_s$, the coherence peaks at positive and negative $E$ can overlap --- producing $\delta\lambda \propto I^2$ scaling beyond a switch-on current $I_0$. Finite temperature results in an exponentially suppressed contribution $\propto I^2$ at small current, as the Fermi function weights states with $\pm v_F p_s$ differently.

Impurities introduce a number of changes to this calculation. The first is a renormalization the parameters of the normal state, shifting for instance the effective $\mathcal{N}_0$. Second, they can smear out the coherence peaks near the gap energy --- altering the temperature dependence for $T>0$, and altering the switch-on of nonlinearity in the current-dependence. Thirdly, impurities which are pair-breaking --- i.e. magnetic impurities in conventional singlet superconductors, or interband scattering impurities in unconventional $s_\pm$ states --- can introduce states inside the gap (depicted schematically as a peak near $E\approx 0$ in Fig.~3C). These state are typically localized for weak isotropic scattering, and so do not shift with $p_s$, and therefore do not contribute to the current/field dependence of $\lambda$.

In order to produce a shift in the penetration depth --- as a function of current or temperature --- impurity states must hybridize to form a weakly dispersive band that does experience a Doppler shift, i.e. the impurities become extended, completely closing the superconducting gap and producing gapless superconductivity. However, the resulting quasiparticles do not disperse linearly, and therefore gives rise to $\delta \lambda \sim I^{2}$ dependence as in the conventional nodeless case; this is borne out in calculations which fully incorporate the effects of impurity scattering on the pairing problem \cite{Sauls.2022}.

By contrast, for a nodal gap, the DOS is $\mathcal{N}(E) = \mathcal{N}_0 E/\Delta$ for $|E|<\Delta$ (Fig.~3D), and the shaded area is a fraction of the large triangle bounded by the dashed line. The area of the large triangle is $\mathcal{N}_0\Delta$, while the small triangle has height $\mathcal{N}_0-\mathcal{N}_0 v_Fp_s/\Delta$ and width $2\Delta - 2v_Fp_s$; the resulting shaded area is $2\mathcal{N}_0v_Fp_s - (\mathcal{N}_0v_F^2/\Delta) p_s|p_s|$. Hence, in addition to the $p_s$ dependence found in fully gapped superconductors, one finds a nonlinear piece that produces \begin{align}
    \lambda^{-2}=2 \mu_0 e^2 \mathcal{N}_0 v_F^2\left(1-\tfrac{v_F}{2 \Delta}\left|p_s\right|\right)
\end{align}
The contribution proportional to $|p_s|$ and hence $|I|$ is the ANLME \cite{Sauls.2022}. At elevated temperatures, this shifts to $ \propto I^2$ dependence. Note that the $\propto |I|$ shift is a correction to $\lambda^{-2}$, whereas our measurement of kinetic inductance probes changes in $\lambda^2$; for small shifts in $\lambda$, one has $\delta \lambda^{2} \propto \delta\lambda^{-2}$.

Note that this contribution is nonanalytic --- the contribution to the supercurrent is proportional to $I|I|$ not $I^2$. A contribution $\propto I^2$ --- corresponding to a $\propto I$ shift in $\lambda^{-2}$ as opposed to $\propto |I|$ --- would correspond to second harmonic generation, and requires the breaking of inversion symmetry, unlike the ANLME.

\subsection*{Circle Fitting}

We extract resonator characteristics from hanger-style transmission measurements by fitting the complex transmission coefficient S$_{21}$(f), measured using a vector network analyzer, to a model that accounts for both ideal and non-ideal effects. The ideal response, defined in terms of the resonator quality factors, is augmented by two key corrections: a global background term \( a\,e^{i(\alpha - 2\pi f \tau)} \), which captures amplitude attenuation, frequency-dependent phase delay, and cable dispersion; and a small asymmetry term \( \phi \) accounting for imperfections in the resonance shape.

The full fitting model is:
\begin{equation}
S_{21}(f) = a\,e^{i(\alpha - 2\pi f \tau)} \left[ 1 - \frac{2(Q_L / Q_C)\,e^{i\phi}}{1 + 2 i Q_L(f / f_r - 1)} \right]
\label{eq:CirFit}
\end{equation}

We use an open-source circle-fitting package to perform the fits in the complex plane \cite{Probst.2015}. The extracted fit parameters from Eqn.~\ref{eq:CirFit} include resonance frequency \( f_r \), loaded quality factor \( Q_L \), and coupling quality factor \( Q_C \), which are used within the main text to quantify the kinetic inductance and applied current using equations \ref{eqn:FreqShift} and \ref{eq:Current} respectively.

\subsection*{Applied Current from Microwave Power}

We use Input-Output theory \cite{gardiner.collett.1985} to infer the current, $I$, flowing through our MoTe$_2$ samples in Fig.\ref{fig:NLME} from the microwave input power that is used to measure the resonators. In quantum optics, Input-Output theory relates the averaged intracavity energy, $\langle E_{\text{cav}}\rangle$ to the microwave input power $P_{\rm{probe}}$, coupling $Q$-factor $Q_c$, and loaded $Q$-factor $Q_l$ for a notch resonator as described below.

We derive Eqn.~\ref{eq:Current} using input-output theory applied to a hanger-type (notch) resonator coupled to a single microwave feedline. In this formalism, the averaged intra-cavity energy $\langle E_{\mathrm{cav}} \rangle$ is related to the complex amplitude $a$ of the resonator mode by
\begin{equation}
|a|^2 = \frac{\langle E_{\mathrm{cav}} \rangle}{\omega_0}, 
\label{eq:a_squared}
\end{equation}
where $\omega_0 = 2\pi f_0$ is the resonant angular frequency.

The normalized input field amplitude $b_{\mathrm{in}}$ is related to the input power by
\begin{equation}
|b_{\mathrm{in}}|^2 = \frac{P_{\mathrm{probe}}}{\omega_0}, \label{eq:b_squared}
\end{equation}
consistent with conventions in quantum optics and cavity quantum electrodynamics.

In steady-state and on resonance, the solution for a driven, underdamped resonator gives the intracavity field amplitude as
\begin{equation}
\label{eq:anb}
|a|^2 = \frac{4 Q_\ell^2}{Q_c} \cdot \frac{|b_{\mathrm{in}}|^2}{\omega_0},
\end{equation}
where $Q_\ell$ is the loaded quality factor and $Q_c$ is the coupling quality factor. Substituting Eqns.~\ref{eq:b_squared} and (\ref{eq:anb}) into Eqn.~\ref{eq:a_squared}, we obtain
\begin{equation}
\langle E_{\mathrm{cav}} \rangle = \omega_0 |a|^2 = \frac{4 Q_\ell^2}{Q_c} \cdot \frac{P_{\mathrm{probe}}}{\omega_0}, 
\label{eq:Ecav}
\end{equation}
as used in Eq.~\ref{eq:Current}. This result is valid for all coupling regimes: under-coupled, critically coupled, and over-coupled, as long as the resonator is driven on resonance and remains in the linear regime.

We note that Eqn.~\ref{eq:IVE} provides a complementary expression connecting the stored energy to the voltage and current in the resonator:
\begin{equation}
|a|^2 = \frac{C |V_{\mathrm{in}}|^2}{2\omega_0} + \frac{L |I|^2}{2\omega_0} = \frac{E}{\omega_0}, \label{eq:IVE}
\end{equation}
which is consistent with the electromagnetic energy stored in an LC oscillator. Together, these relations provide a complete connection between microwave input power, resonator field amplitudes, and physical quantities such as current.

Substituting Eqn.~\ref{eq:Ecav} into the identity $\langle E_{\mathrm{cav}} \rangle = \tfrac{1}{2} L_0 I^2$, we obtain the expression for current:
\begin{equation}
I = \sqrt{ \frac{8 Q_\ell^2}{\omega_0 L_0 Q_c} P_{\mathrm{probe}} }
\label{eq:Current}
\end{equation}
which is used to estimate the peak current flowing through the current antinode of the inductive element of the resonator.

\newpage

\widetext

\section*{Supplementary Information}

\subsection{Analysis of multiple devices}

\begin{table}[h]
\begin{tabular}{ l c c c } 
\hline
sample~~~~~&~~~~width~($\mu$m)~~~~&~~~~length ($\mu$m)~~~~&~~~~thickness (nm)~~~~\\
\hline
KI18 &  $7.9 \pm 0.8$  & $47 \pm 0.5$ & $89 \pm 0.4$ \\
KI22 & $16.9 \pm 1.7$ & $151.4 \pm 1.5$ & $200 \pm 0.7$ \\
KI23 &  $9.7 \pm 1.0$  & $42.5 \pm 0.4$ & $205 \pm 0.6$ \\
KI26 ($b$-axis) & $9.3 \pm 0.9$ & $15.2 \pm 0.2$ & $82 \pm 3.2$ \\
KI31 &  $8.6 \pm 0.9$  & $29.8 \pm 0.3$ & $289.1 \pm 0.6$ \\
\hline
\end{tabular}
\caption{\textbf{Dimensions of MoTe$_2$ samples.} }
\label{tab:SampleDimensions}
\end{table}

\begin{table}[h]
\begin{tabular}{ l c c c } 
\hline
sample~~~~~&~~~~n~\\
\hline
KI18 &  $1.99 \pm 0.04$  \\
KI22 & $1.97 \pm 0.16$ \\
KI23 &  $1.97 \pm 0.06$ \\
KI26 ($b$-axis) & $1.84 \pm 0.18$ \\
\hline
\end{tabular}
\caption{\textbf{Power law of temperature dependence in MoTe$_2$ samples.} }
\label{tab:TdepPower}
\end{table}

\begin{figure}[h]
\centering
\includegraphics[width=0.45\columnwidth]{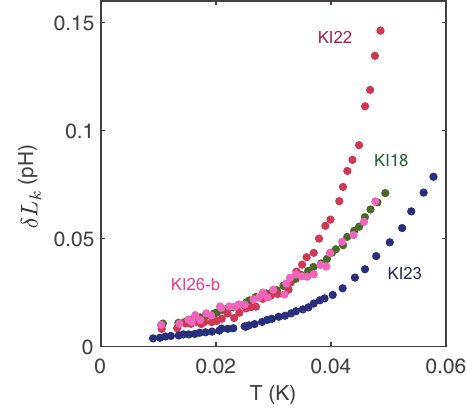} 
\caption{\textbf{Temperature dependent kinetic inductance in MoTe$_2$ devices.} Fine-temperature measurements display reproducible non-saturating kinetic inductance at low temperatures $T < 0.3T_c$.}
\label{fig:Tdep all samples}
\end{figure}

The temperature dependence of the resonance frequency provides direct information about the kinetic inductance and, therefore, the penetration depth of the superconductor. As shown in Fig.~\ref{fig:Tdep all samples}, we observe a monotonic increase in resonance frequency as the temperature decreases, consistent with the reduction in kinetic inductance expected for a superconducting system. Importantly, we find no evidence of saturation down to 9~mK, indicating that the superfluid stiffness continues to grow as thermal excitations diminish. We compare the temperature dependence across samples and report power law scaling of approximately $n = 2$ (see Table~ \ref{tab:TdepPower}). In all cases, the extracted $\delta f_\mathrm{res}(T)$ deviates strongly from exponential behavior and instead follows a power law, consistent with nodal superconductivity.

\begin{figure}[hbt!]
\centering
\includegraphics[width=0.45\columnwidth]{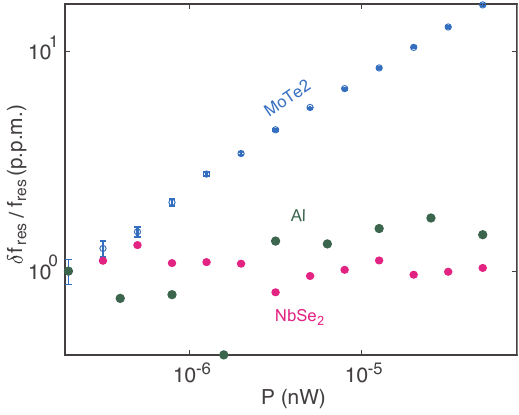} 
\caption{\textbf{Normalized frequency shift of MoTe$_2$ compared to NbSe$_2$ and Al at 9~mK.} Among these materials, only MoTe$_{2}$ exhibits $\sim |I| \propto \sqrt{P}$ dependence in accordance with the anomalous nonlinear Meissner effect as described in the main text.}
\label{fig:MoTe2vAlNbSe2}
\end{figure}
Figure~\ref{fig:MoTe2vAlNbSe2} shows the normalized fractional frequency shift as a function of input microwave power for MoTe$_2$, NbSe$_2$, and aluminum at a temperature of 9~mK. While the aluminum and NbSe$_2$ devices show negligible frequency dependence across the measured power range, MoTe$_2$ exhibits a clear monotonic increase in frequency shift beginning at sub-pico-watt powers. This power-dependent shift in resonance frequency is a hallmark of the nonlinear Meissner effect, and its onset at significantly low excitation powers in MoTe$_2$ suggests a strongly enhanced current sensitivity of the superfluid response, consistent with a nodal order parameter. The absence of a similar effect in the control materials under identical measurement conditions further supports its intrinsic origin in the MoTe$_2$ superconductor.

\begin{figure}[t]
\centering
\includegraphics[width=\columnwidth]{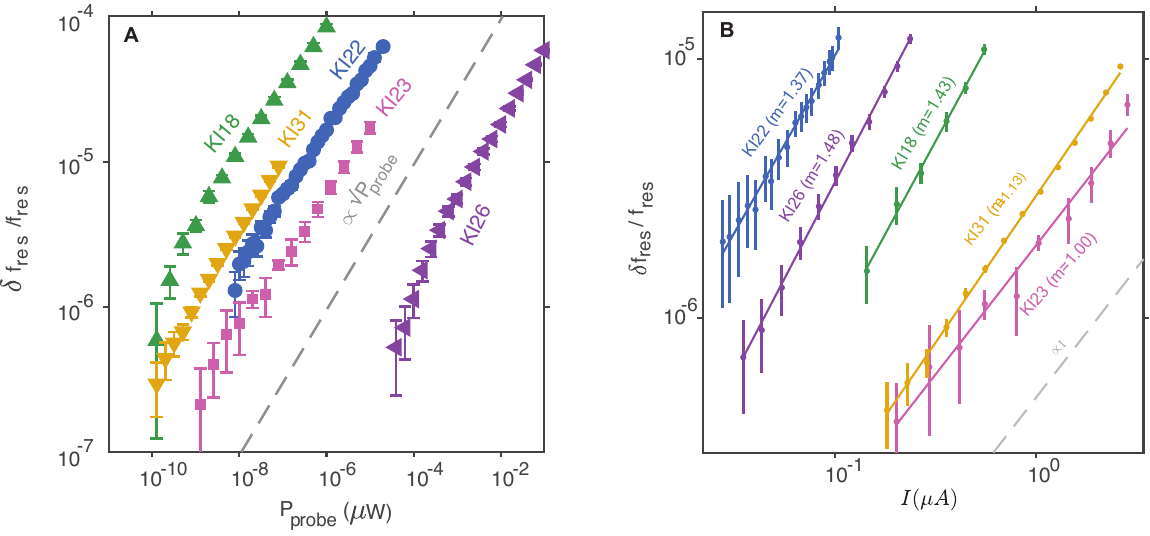} 
\caption{\textbf{Frequency response as a function of applied microwave power and inferred microwave current measured in five MoTe$_2$ devices.} Normalized resonance frequency shifts, (A) $\delta f/f = |f(T,P)-f(T,P_0)|/f(T,P_0)$ and (B) $\delta f/f = |f(T,I)-f(T,I_0)|/f(T,I_0)$ are plotted as a function of (A) the applied microwave power $P_{probe}$ and (B) its corresponding current $I$, inferred using input-output theory (see Methods). $P_0$ and $I_0$ denote the lowest applied microwave power at each temperature $T$. (B) Fitted exponents $m$ are indicated for each sample. Error bars represent the 95 percent confidence intervals from the extracted resonance frequency. The dashed gray lines are a guide to the eye illustrating square root and linear power-law behavior as a function of power and current respectively. Data and fits are restricted to powers well below the onset of higher-order nonlinearities.}
\label{fig:PowerDep_Base}
\end{figure}

We then use Eqn.~\ref{eq:Current} to plot the fractional frequency change due to the microwave input power as a function of converted peak current at the current antinode for multiple MoTe$_2$ devices to show the repeatability and robustness of this effect across various samples, as shown in Fig.~\ref{fig:PowerDep_Base}.

\subsection{Details of the Temperature-Dependent Data Fitting Procedure}

In determining the low-temperature scaling of the superfluid stiffness, we must estimate the zero temperature value of the resonance frequency. Our procedure is to determine a polynomial of best fit, and then arrive at an estimate by extrapolation to zero temperature; we confirm that the scaling behavior of the stiffness remains robust to modest changes in this estimate. 

\begin{figure}[t]
\centering
\includegraphics[width=0.65\columnwidth]{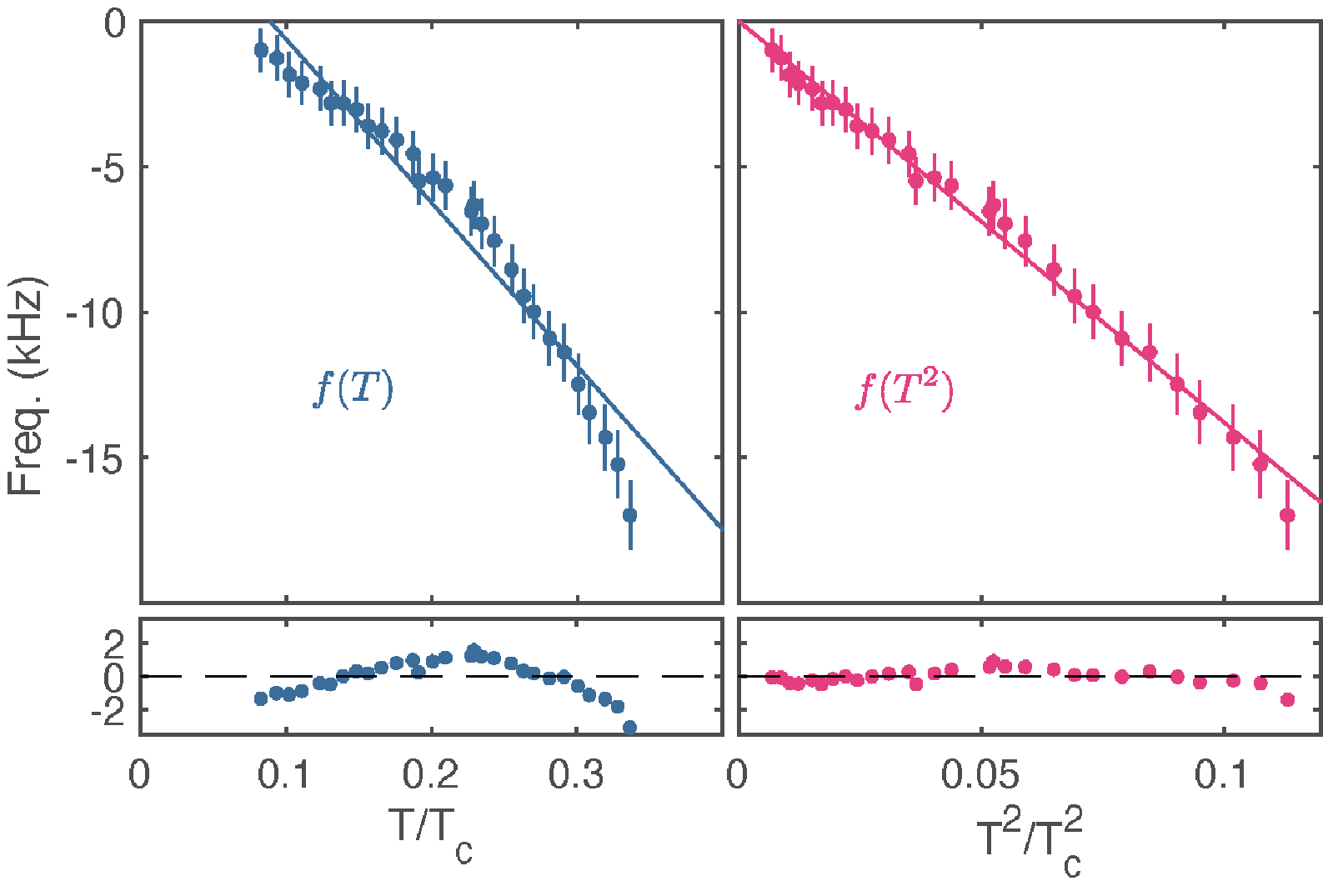} 
\caption{\textbf{Comparing power-law fits of the temperature dependent frequency.} The resonance frequency for $T<T_c/3$ is fit with a functional form $f(T)$ which is either a linear (Left) or quadratic (Right) polynomial. The linear fitting is plotted as a function of $T/T_c$, whereas the quadratic fitting is plotted as a function of $(T/T_c)^2$, so that in both cases the data are compared with a line of best fit. Crucially, fitting the data to linear scaling results in a systematic drift of the data away from the line of best fit at lower and larger temperatures, as clearly illustrated by the residual plot (bottom); such a drift does not occur for the quadratic fitting, for which the residuals exhibit no trend as a function of temperature. Similar systematic drift occurs when fitting the data to exponential dependence.}
\label{fig:fit_comp}
\end{figure}

For comparison, in Fig.~\ref{fig:fit_comp} we show the result of fitting the data of main text Fig.~2 with a linear and quadratic polynomial. Fitting the data to linear scaling, one observes a systematic drift of the data away from the fit as a function of temperature, with residual errors that steadily increase at lower temperatures. By comparison, the residual error remains small and approximately temperature independent for the quadratic fit, illustrating strong evidence of $\sim T^2$ scaling behavior. Exponential fitting would produce an even more pronounced systematic drift of the error as a function of temperature.

Having determined that the quadratic fit accurately describes the data, we extrapolate to zero temperature to determine an estimate of the zero temperature frequency. Then, we return to the raw data and subtract this zero temperature value, and examine the data on a log-log scale to illustrate power-law behavior. We fit the data on a log-log scale using linear regression to determine the power law of best fit; the result is $\lambda^2(0) - \lambda^2(T) \propto T^n$ with an exponent of $n=2.12\pm 0.15$ for the data in Fig.~2, as cited in the main text.

We stress it is essential to examine the data on a log-log scale to accurately assess the low-temperature scaling in a rigorous manner, rather than simply plotting the data in absolute units and arbitrarily fitting theoretical functional forms, as done in many prior studies of temperature dependent superfluid stiffness. It is also essential to have many data points in the regime $T<T_c/3$ to confidently exclude or confirm the presence of gap nodes; an advantage of our experimental technique over many prior works is that we have sufficiently low-temperature resolution to accurately probe this regime for a low $T_c$ superconductor.

To determine the scaling behavior with applied current, it is important to exclude data taken at the lowest power. The reason for this is that the technique becomes increasingly less reliable in this regime, highlighted by the systematically increasing error bars at small $I$ in Fig.~\ref{fig:NLME}. To determine the slope of the data on a logarithmic scale, we employ a statistical test for the exclusion of data with larger errors: we exclude data points $(\log I_n, \log \delta f_n)$ for which the error bar $\Delta (\log \delta f_n)$ is larger than the difference in magnitude between adjacent measurements, i.e. $\Delta (\log \delta f_n) > \log \delta f_{n+1}-\log \delta f_n$. Such data have less statistical power in determining the gradient, and should naturally be excluded when fitting the slope. After excluding the unreliable data taken at low power, we then fit the data using polynomial regression on a logarithmic scale, as done for the temperature dependence.

\newpage
\subsection{Harmonic Analysis of the Anomalous nonlinear Meissner Effect}
While Eqn.~\ref{eqn:FreqShift} expresses the resonance frequency shift due to the change of kinetic inductance, the anomalous nonlinear Meissner effect (ANLME) depends specifically on the magnitude of the microwave current. To capture this, we express the kinetic inductance due to the ANLME as \begin{equation}
    L_{\text{ANLME}} = L_{K}\left(1+\frac{|I_{mw}(t)|}{I^*}\right)
\end{equation}\label{eq:Lk(I)}
where $I^*$ is a characteristic current comparable to the critical current. When $I=I^*$, $p_sv_F\simeq\Delta$. We describe the absolute value of the time-varying microwave current as:
\begin{eqnarray}
    \left|I_{mw}(t)\right| &=& I_{mw}|\sin{2\pi f_{\text{res}}t}|
\end{eqnarray}
To analyze this using harmonic components, we approximate $\left|\sin{2\pi f_{\text{res}}t}\right|$ using a Fourier series expansion. This yields
\begin{eqnarray}
    \left|I_{mw}(t)\right|&\simeq& I_{mw}\left[\frac{2}{\pi}-\frac{4}{\pi}\left(\frac{1}{3}\cos{4\pi f_{\rm{res}}t}+\frac{1}{15}\cos{8\pi f_{\rm{res}}t}\right)\right]
\end{eqnarray}
This decomposition highlights that the nonlinear response of the kinetic inductance includes both a DC component $2I_{mw}/\pi$ and higher harmonics at or above $2f_{\rm{res}}$, which are neglected because it is the DC term which results in the time-averaged shift in the kinetic inductance and, hence, the resonance frequency observed in Fig.~3 of the main text.

We emphasize one critical nuance: that the correction to the kinetic inductance from the ANLME is not simply linear in the current, but rather takes a non-analytic form which depends on the absolute value of the current. The nonanalyticity of this contribution means that it cannot be derived from a perturbative linear response theory --- a subtlety that was appreciated early in the literature \cite{Yip.1992, Sauls.2022}. Moreover, a linear current dependence may only arise in the presence of inversion symmetry-breaking, whereas the ANLME arises generically in nodal superconductors and does not require inversion symmetry-breaking.

\newpage
\subsection{Temperature Dependence of Stiffness in Additional Devices}
\begin{figure}[htb!]
    \centering
    \includegraphics[width=\columnwidth]{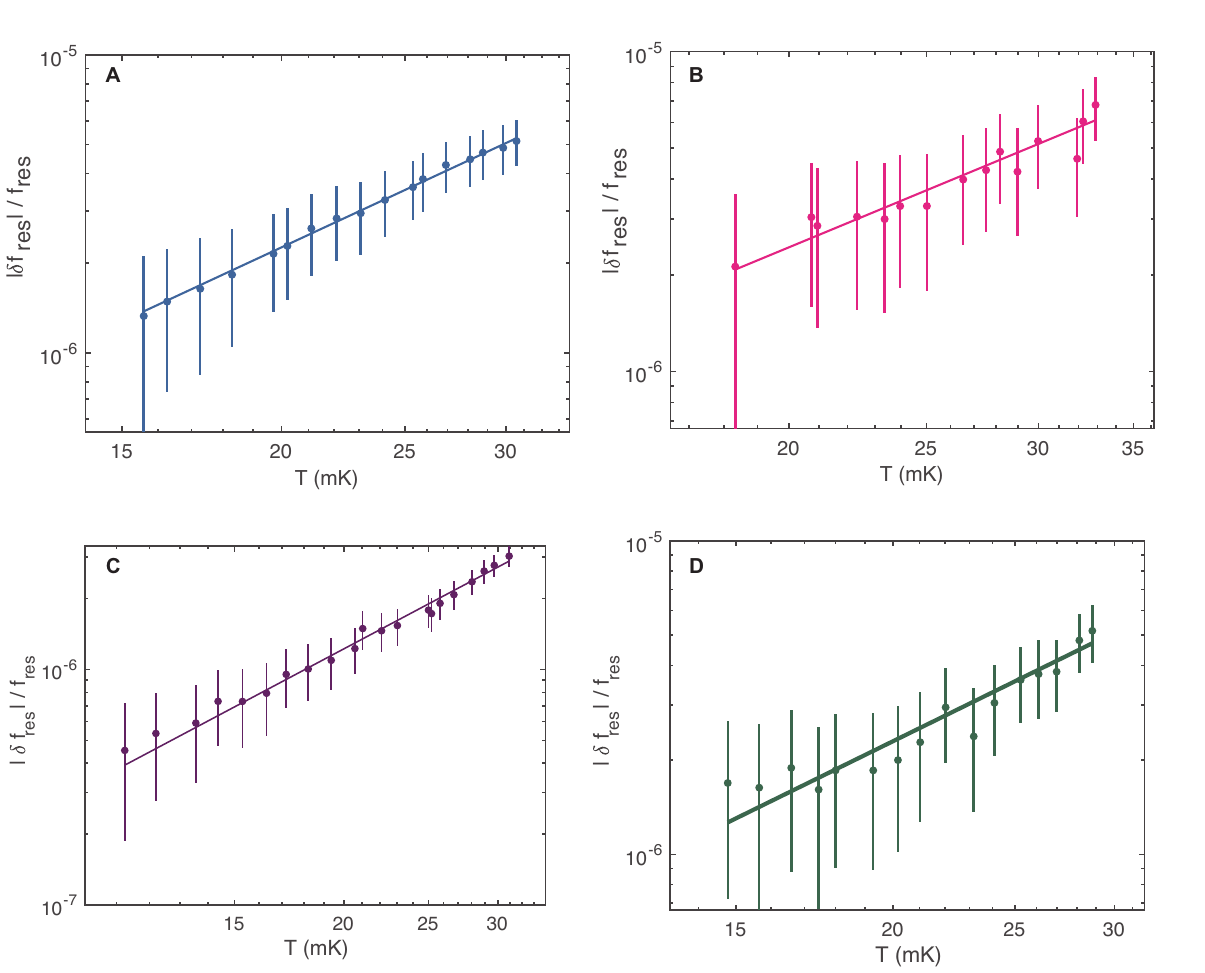} 
    \vspace{1cm}
    \caption{\textbf{Determining reproducibility of $T^2$ behavior of the low-temperature penetration depth in several devices.} Linear fitting of the shift in frequency versus temperature on a log-log scale, using the data analysis methods described in  Supplementary Information D. Panel (B) shows data from the single $b$-axis sample. (A-D) are oriented along the crystalline $a$-axis. Fitted values for $n$ in $T^n$ are reported in Table~\ref{tab:TdepPower}.}
    \label{fig:extra devices}
    \end{figure}
    
We apply the previously described fitting technique to examine the temperature dependence in four MoTe$_2$ devices: KI18, KI22, KI23, and KI26b, illustrating the reproducibility of the $\sim T^2$ scaling behavior; the results are shown in Fig.~\ref{fig:extra devices}. The results are consistent with $\sim T^2$ scaling: for KI18 we find $n = 1.99 \pm 0.04$, for KI22 $n = 1.97 \pm 0.16$, KI23 $ n = 1.97 \pm 0.06$, and for KI26b $n = 1.84 \pm 0.18$.

\newpage
\bibliography{refs}

\end{document}